\documentclass[DIV12]{scrartcl}

\usepackage{hyperref}
\usepackage[utf8]{inputenc}
\usepackage{tabularx}
\usepackage{setspace}
\usepackage{graphicx}
\usepackage{booktabs}
\usepackage{amsmath}

\hypersetup{
	colorlinks=true,
	linkcolor=black,
	citecolor=black,
	urlcolor=black
}

\date{}
\usepackage{authblk}
\usepackage{blindtext}

\usepackage{natbib}
\title{Talking Drums: Generating drum grooves with neural networks}

\usepackage{fancyhdr}
\begin{document}

\author[1]{P. Hutchings\thanks{pehut2@student.monash.edu}}

\affil[1]{\small Monash University, Melbourne, Australia}

\maketitle

\thispagestyle{fancy} 

\begin{abstract}
Presented is a method of generating a full drum kit part for a provided kick-drum sequence.  A sequence to sequence neural network model used in natural language translation was adopted to encode multiple musical styles and an online survey was developed to test different techniques for sampling the output of the softmax function.  The strongest results were found using a sampling technique that drew from the three most probable outputs at each subdivision of the drum pattern but the consistency of output was found to be heavily dependent on style.

\bigskip

\noindent {\textbf{Keywords:}} RNN, percussion, generative music, translation

\end{abstract}

\section{Introduction}
This research details the development of a percussion-role agent as part of a larger project where virtual, self-rating agents with different musical roles work in a process of co-agency to generate music compositions in real-time \citep{hutchings2017using}.  The percussion-role agent was developed for generating multiple possible multi-instrument percussion parts to accompany provided melodies and harmonies in real-time.

A neural network based agent was developed to incorporate a range of different music styles from a large corpus of compositions and to utilise a softmax function as part of the self-rating process.  A network architecture used in natural language translation was adopted based on the idea that a percussion score could be considered as containing multiple drums `speaking' different languages but saying the same thing at the same time.  The network was trained on a collection of drum kit scores from over 250 pop, rock, funk and Afro-Cuban style compositions and patterns from drum technique books.  The output of the network was evaluated from an online survey and a physical interface was developed for feeding kick-drum parts into the network.

\subsection{Related work}
Markov models \citep{hawryshkewich2010beatback} \citep{tidemann2008drum}, generative grammars \citep{bell1992bol} and neural network models \cite{DBLP:journals/corr/ChoiFS16} have all been shown to be effective in the area of drum score generation.  The approach shown in this paper is based on the requirements of generating an agent for a multi-agent composition system.  Research in this area has demonstrated the need for agent models to match the needs of the overall system \citep{eigenfeldt2009realtime}.

The similarities and differences between music and natural language have been explored in detail \citep{patel2003language} \citep{mithen2011singing}.  While distinct differences exist in terms of cognitive processing, semantics and cultural function, there are similarities in the structure of phrases that have lead to the use of natural language processing techniques in the analysis and generation of music.

\subsection{Translation model}

Generating a full drum kit score based on the rhythm of one or more individual instruments in the kit is a problem with different challenges than natural language translation.  All translations are one to one in word count.  Music is a non-semantic form of communication which allows for and values greater structural variation than spoken language so imperfect translations can still be effective.  Conversely because there is no perfect translation, there are many different outputs for a given input in the training data, decreasing convergence during training.  The problem can also be viewed as one of data-expansion as a single instrument part is expanded to fill a full drum kit with multiple concurrent instruments being used. To take advantage of these strengths and diminish the weaknesses of a translation based neural network model a new syntax for expressing drum parts was developed.

\section{Method}

\subsection{Data preprocessing}
A collection of 250 drum kit scores in 4/4 were found on drum tablature websites and books and parsed into a music-XML format.  Tracks were selected based on the most viewed web-pages for rock, pop, funk and Afro-Cuban styles of music and were each checked for accuracy by comparing with the original recordings by ear.  Pop, rock and funk styles were selected due to their global popularity and typical use of a standard drum kit.  The Afro-Cuban style was added to this list to see if some of the stricter idiomatic structures of the style, such as the `clave' rhythmic pattern, could be preserved.  Afro-Cuban and funk drum tablatures were more difficult to find so the tablatures were augmented with patterns from drum technique instruction books.  For each genre a total of 7000-7500 bars were parsed.

Each bar was divided into 48 subdivisions, allowing all triplet and tuple divisions down to the resolution of semiquaver triplets to be represented.  Each division was given a word token that represented the drums being hit on that subdivision and barlines were replaced with a word token describing the musical style which allowed multiple styles to be encoded in a single RNN network.

The tokenised phrase in Equation~\ref{eq1} represents a kick-drum being kicked on each beat of a single 4/4 bar and a `pop' style description.  

\begin{equation} \label{eq1}
pop\ K \ o \ o \ o \ o \ o \ o \ o \ o \ o \ o \ o \ K \ o \ o \ o \ o \ o \ o \ o \ o \ o \ o \ o \ K \ o \ o \ o \ o \ o \ o \ o \ o \ o \ o \ o \ K \ o \ o \ o \ o \ o \ o \ o \ o \ o \ o \ o
\end{equation}

The full list of letter representations used to create word tokens are presented in Table~\ref{tab:tokens}.  Composition segments of 4 bars were used as sentences for training the neural network with kick-drum patterns used as inputs to the encoder layer and the rest of the drum parts in the decoder layer.  Encoder input sequences were reversed and encoded using one-hot encoding. The kick-drum was selected as the input language because it is usually used to mark the beat of a composition and small changes can dramatically affect the feeling of time.

\begin{table}[h]
\centering
\caption{Letter representations of drums}
\label{tab:tokens}
\begin{tabular}{|c|c|c|c|c|c|c|c|c|}
\hline
Drum   & Cymbal & Hi-hat & Snare & High Tom & Tom & Floor Tom & Kick & None \\ \hline
Letter & C      & H      & S     & T        & t   & F         & K    & o    \\ \hline
\end{tabular}
\end{table}

\subsection{Network architecture}
The neural network has an RNN sequence-to-sequence architecture \citep{sutskever2014sequence} using the Tensorflow deep-learning framework \citep{tensorflow2015-whitepaper}.  A model layer of size 128 and 3 layers produced a perplexity of 1.15 when trained with a learning rate of 0.55 and a gradient descent optimiser.  This was the lowest perplexity achieved from a manual testing of variations to these hyper-parameters.  Hidden states were initialised with all zero values and updated at each step of training.

\section{Evaluation}
An online survey was generated to find a sampling technique that human listeners found preferable.  The survey was advertised on social media groups related to drumming and computer music and run for two weeks.

\subsection{Survey}
\begin{figure}[h]
\centering
\includegraphics[width=.7\textwidth]{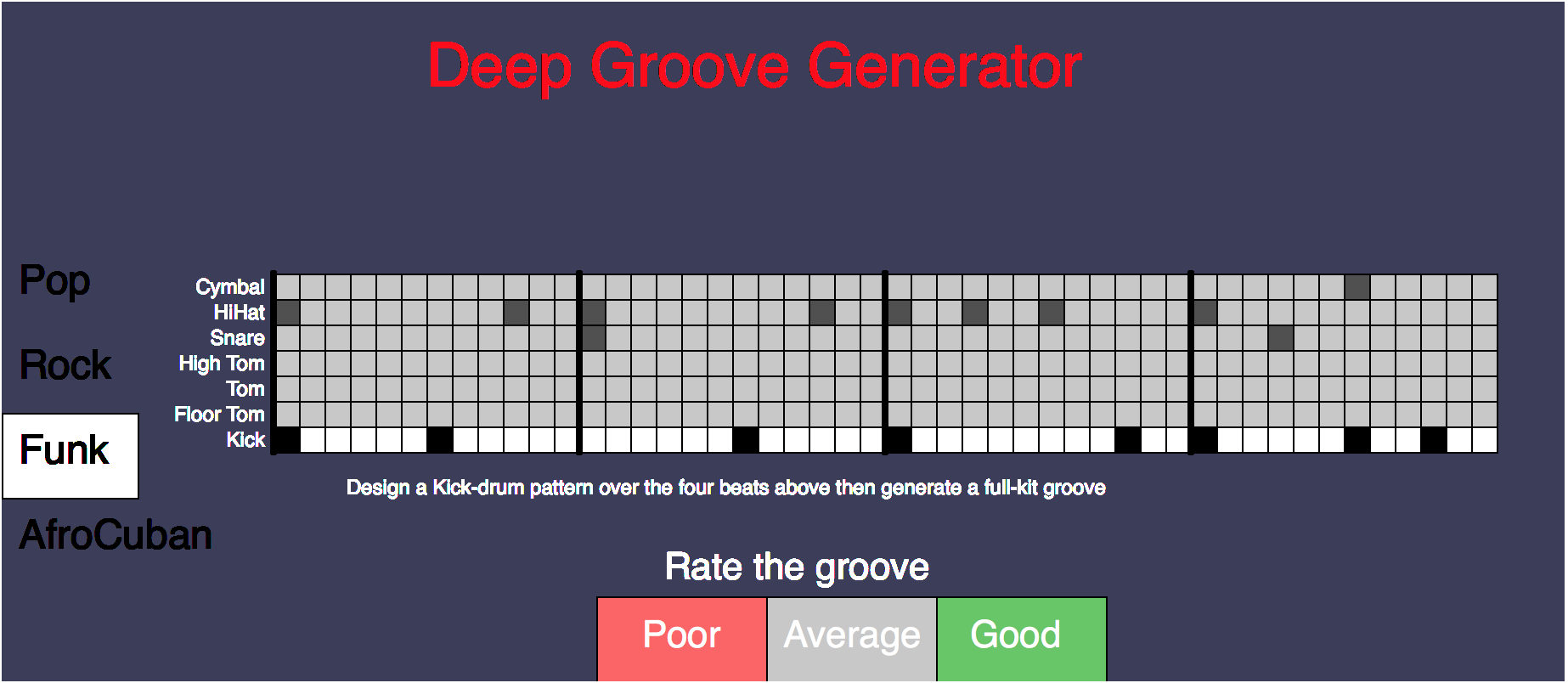}
\caption{Interface for the online evaluation survey}
\label{fig:example}
\end{figure}

Participants were presented with a style menu and a 48 step sequence with an editable kick-drum line that they could use to design a four beat kick-drum pattern as seen in Fig.~\ref{fig:example}.  After clicking a `Generate Groove' button on the interface, the other instrument parts would be generated and a loop of the pattern would begin playing with sounds sampled from drum kits.  Participants were then asked to rate the groove as poor, average or good.  The survey was designed to encourage a fast and playful experience, so demographic data was not asked or collected.

Each time a groove was generated the web application ran the input through the neural network and randomly selected a sampling method.  Three sampling methods were tested: A greedy decoder (Method 1), a roulette-wheel sampler across all probabilities (Method 2) and a roulette-wheel sampler of the three most probably tokens at each subdivision (Method 3).

\subsection{Results}
A total of 1278 groove evaluations were recorded in the survey.

\begin{table}[h]
\centering
\caption{Survey results for different sampling methods}
\label{tab:results}
\begin{tabular}{l|l|l|l|l|l|l|}
\cline{2-7}
                               & \multicolumn{3}{c|}{Raw} & \multicolumn{3}{c|}{Normalised} \\ \cline{2-7} 
                               & Good  & Average  & Poor  & Good     & Average    & Poor    \\ \hline
\multicolumn{1}{|l|}{Method 1} & 91    & 276      & 30    & 0.23     & 0.70       & 0.08    \\ \hline
\multicolumn{1}{|l|}{Method 2} & 100   & 217      & 125    & 0.23     & 0.49       & 0.28    \\ \hline
\multicolumn{1}{|l|}{Method 3} & 172   & 183      & 84    & 0.39     & 0.42       & 0.19    \\ \hline
\end{tabular}
\end{table}

As shown in Table~\ref{tab:results} the model produced full drum-kick patterns that were deemed to be average or good in a majority of ratings on the web survey.  Of the three sampling methods it can be observed that the greedy encoder had a tendency towards results that participants deemed average.  The roulette wheel sampling used in Method 2 had the highest rate of `poor' ratings.  Overall the best performer was the sampler that drew from the three most probable tokens at each subdivision.  Examples of 5 drum patterns for each sampling method are available to listen to at \href{https://doi.org/10.6084/m9.figshare.4903181.v1}{https://doi.org/10.6084/m9.figshare.4903181.v1}. 

\begin{table}[h]
\centering
\caption{Mean rating for mean initial probabilities of selected notes.}
\label{tab:prob}
Poor =0, average =1, good = 2
\begin{tabular}{|l|l|l|l|l|l|l|l|}
\hline
Mean probability & 0.2-0.3 & 0.3-0.4 & 0.4-0.5 & 0.5-0.6 & 0.6-0.7 & 0.7-0.8 & 0.8-0.9 \\ \hline
Mean rating      & 0.25    & 0.27    & 0.58    & 1.14    & 1.32    & 1.54    & 1.22    \\ \hline
\end{tabular}
\end{table}

\section{Discussion and future work}
The ratings in Table~\ref{tab:prob} peaked when the average probability was between 0.7-0.8, below the maximum observed bracket of 0.8-0.9.  This may be a result of participants valuing familiar but different drum patterns over patterns that they may have heard in songs they know.  The significantly higher rating of one band of probability range supports the use of the model in the intended application of a multi-agent system as it provides a means of self-rating output.  Mean ratings of Afro-Cuban style patterns were significantly lower (24\% poor) than for other styles (16-18\% poor) which may be the result of stylistic bias of the participants or could suggest important elements of the style are not represented in the model output.

A syntax for expressing desired accents is being developed as an encoder to expand the pallet and may improve results in the Afro-Cuban and other styles.  A physical drum-pedal interface has been developed to test the system with drummers in a natural playing position.

\bibliography{paper}

\end{document}